\documentclass[showpacs, showkeys,twocolumn,epsf]{revtex4}
\usepackage{graphicx}
\begin{document}
\newcommand{\dfrac}[2]{\frac{\displaystyle #1}{\displaystyle #2}}
\title{ Electrical Neutrality and Symmetry Restoring
Phase Transitions at High Density in a Two-Flavor
Nambu-Jona-Lasinio Model\footnote{The project supported by the
National Natural Science Foundation of China under
Grant No.10475113.} \\
}
\author{WANG Xiao-Ming and ZHOU Bang-Rong}
\affiliation{College of Physical Sciences,
Graduate School of the Chinese Academy of Sciences, Beijing
100049, China}
\date{}
\begin{abstract}
A general research on chiral symmetry restoring phase transitions
at zero temperature and finite chemical potentials under
electrical neutrality condition has been conducted in a
Nambu-Jona-Lasinio model to describe two-flavor normal quark
matter. Depending on that $m_0/\Lambda$, the ratio of dynamical
quark mass in vacuum and the 3D momentum cutoff in the loop
integrals, is less or greater than 0.413, the phase transition
will be second or first order. A complete phase diagram of $u$
quark chemical potential versus $m_0$ is given. With the
electrical neutrality constraint, the region where second order
phase transition happens will be wider than the one without
electrical neutrality limitation. The results also show that, for
the value of $m_0/\Lambda$ from QCD phenomenology, the phase
transition must be first order.
\end{abstract}
\pacs{11.10.Wx; 11.30.Rd; 11.10.Lm; 11.15.Pg} \keywords{Normal
quark matter, electrical neutrality, Nambu-Jona-Lasinio model,
high density chiral symmetry restoring, first and second order
phase transition} \maketitle
\section{Introduction}\label{Intro}
Nambu-Jona-Lasinio (NJL) model \cite{kn:1} is a good low energy
phenomenological model which can be used to simulate Quantum
Chromodynamics (QCD) and is suitable for researching spontaneous
breaking of symmetry and its restoring at finite temperature and
finite chemical potential in quark matter\cite{kn:2,kn:3}. In a
realistic research of the normal quark matter without color
superconductivity, owing to that the quarks carry electrical
charges,  one must impose electrical neutrality condition on the
quark matter. In this paper, we will generally examine symmetry
restoring phase transitions at zero temperature and high density
in a NJL model with two-flavor normal quark matter under
electrical neutrality condition, however, the confinement problem
which is beyond the power of a NJL model will not be touched on.
This examination will make us get more deep-going understanding of
the affect of electrical neutrality on phase transitions and
certainly has important theoretical significance. It is noted that
such general research has not yet appeared in present literature.
The preceding relevant researches either had completely not
touched on electrical neutrality requirement \cite{kn:4} or only
limited relevant parameters e.g. the momentum cutoff $\Lambda$ and
the dynamical quark mass $m_0$
in vacuum to some given values by phenomenology.\cite{kn:5}.\\
\indent It is a well-known fact that in the vacuum of QCD, chiral
symmetry is inevitably spontaneously broken
\cite{kn:1,kn:6,kn:7,kn:8}, so in our model, we will first assume
that the quarks in vacuum have acquired a common non-zero
dynamical mass from the quark-antiquark condensates, then research
that, as increase of the quark chemical potential, the dynamical
quark mass could change to zero and the chiral symmetry will
finally be restored . In the process, the electrical neutrality
condition will always be maintained and the order of the symmetry
restoring phase transitions will be the key point of our concern.\\
\indent For keeping electrical neutrality of the quark matter, we
will add the contribution of free electron gas to the effective
potential. The discussions will be made in the mean-field
approximation. Throughout the paper,a three-dimension momentum
cutoff will be used.\\
\indent The paper is arranged as follows. In Sect.\ref{eff} we
will give the effective potential of a NJL model describing
two-flavor normal quark matter, its extreme value equation and the
electrical neutrality condition. In Sect.\ref{small}, we will
discuss the second order phase transition in small dynamical quark
mass in vacuum, or equivalently, in large momentum cutoff, and in
Sect.\ref{large} expound the second and the first order phase
transition in larger ratio of the above two parameters. Finally,
in Sect.\ref{conc} we come to our conclusions.
\section{Effective potential,gap equation and electrical
neutrality condition}\label{eff} The Lagrangian of the NJL model
describing the two-flavor normal quark matter can be expressed by
\begin{eqnarray}
{\cal L}&=&\bar{q}i\gamma^{\mu}\partial_{\mu}q
+G_S[(\bar{q}q)^2+(\bar{q}i\gamma_5\vec{\tau}q)^2]
\end{eqnarray}
with the quark Dirac fields $q$ in the $SU_f(2)$ doublet and the
$SU_c(3)$ triplets, i.e.
$$q=\left(
\begin{array}{c}
  u_i\\
  d_i \\
\end{array}
\right)\;\; i=r,g,b,
$$
where the subscripts $i=r, g, b$ denote the three colors (red,
green and blue) of quarks, $\vec{\tau}=(\tau_1,\tau_2,\tau_3)$ are
the Pauli matrices, $G_S$ is the four-fermion coupling constants
and we have omitted the bare mass of the quarks. Assume that the
four-fermion interactions can lead to the scalar quark-antiquark
condensates $\langle\bar{q}q\rangle=\phi$, then the chiral
$SU_{fL}(2)\otimes SU_{fR}(2)$ flavor symmetry of the Lagrangian
(1) will be spontaneously broken down to $SU_{fV}(2)$ and the
quarks will get dynamical mass $m=-2G_S\langle\bar{q}q\rangle$. In
the mean field approximation \cite{kn:9}, we can write the
effective potential of the model in the $T\rightarrow 0$ limit by
\begin{eqnarray}
&&V(m,\mu,\mu_e) =\nonumber\\
&&\frac{m^2}{4G_S}-6\int\frac{d^3p}{(2\pi)^3}\left\{
2(E_p-p)\right.\nonumber\\
&&\left.+\left[\theta(\mu_u-E_p)(\mu_u-E_p)+ (\mu_u\rightarrow
\mu_d)\right]\right\}-\frac{\mu_e^4}{12\pi^2}
\end{eqnarray}
where $E_p=\sqrt{\vec{p}^2+m^2}$, $\mu=-\partial V/\partial n$ is
the quark chemical potential corresponding to the total quark
number density $n$, $\mu_e$ is the chemical potential of electron
and
\begin{equation}
\mu_u=\mu-\frac{2}{3}\mu_e, \;
\mu_d=\mu+\frac{1}{3}\mu_e=\mu_u+\mu_e
\end{equation}
are respectively the chemical potentials of the $u$ and $d$
quarks. The second equality in Eq.(3) is usually refers as beta
equilibrium \cite{kn:9}. In vacuum we have $\mu=\mu_e=0$, thus the
effective potential (2) is reduced to
$$
V_0(m)=\frac{m^2}{4G_S}-12\int\frac{d^3p}{(2\pi)^3}(E_p-p).
$$
With a 3D momentum cutoff $\Lambda$, we may find out that the
extreme value points of $V_0$ are 1) $m=0$ and 2) $m=m_0$, where
$m_0$ obeys the gap equation
\begin{equation}
\frac{1}{2G_S}=\frac{3}{\pi^2}\left(\Lambda\sqrt{\Lambda^2+m_0^2}
-m_0^2\ln\frac{\Lambda+\sqrt{\Lambda^2+m_0^2}}{m_0}\right).
\end{equation}
It is easy to verify that if
\begin{equation}
1/2G_S<3\Lambda^2/\pi^2,
\end{equation}
then $m=0$ will be a maximum point and simultaneously Eq.(4) will
have non-zero solution $m_0$ which is a minimum point. This means
spontaneous breaking of chiral symmetry in vacuum. This will
assumedly be our presupposition of discussions in this paper.
Hence we may replace $1/2G_S$ in the effective potential
$V(m,\mu,\mu_e)$ by using Eq.(4). As a result, we obtain from
Eq.(2)
\begin{widetext}
\begin{eqnarray}
&&V(m,\mu,\mu_e)=\nonumber\\
&&\frac{3}{4\pi^2}\left\{ m^2\left(
2\Lambda\sqrt{\Lambda^2+m_0^2}-2m_0^2
\ln\frac{\Lambda+\sqrt{\Lambda^2+m_0^2}}{m_0}\right)
+2\Lambda^4-\Lambda\sqrt{\Lambda^2+m^2}(2\Lambda^2+m^2) +m^4
\ln\frac{\Lambda+\sqrt{\Lambda^2+m^2}}{m}\right\}\nonumber\\
&&+\frac{3}{4\pi^2}\left\{\theta(\mu_u-m)\left[
\frac{m^2\mu_u\sqrt{\mu_u^2-m^2}}{2}
-\frac{\mu_u(\mu_u^2-m^2)^{3/2}}{3}
-\frac{m^4}{2}\ln\frac{\mu_u+\sqrt{\mu_u^2-m^2}}{m}\right]
+(\mu_u\rightarrow\mu_d)\right\}-\frac{\mu_e^4}{12\pi^2}
\end{eqnarray}
\end{widetext}
For deriving electrical neutrality condition, it is noted that the
electrical charge density in the two-flavor quark matter with
electrons is
$$n_Q=\frac{2}{3}n_u-\frac{1}{3}n_d-n_e,
$$
from which we may obtain
$$
\mu_e=-\frac{\partial V}{\partial n_e}=-\frac{\partial V}{\partial
n_Q}\frac{\partial n_Q}{\partial n_e}=-\mu_Q.
$$
Hence the electrical neutrality condition will become
$n_Q=-\partial V/\partial \mu_Q=\partial V/\partial\mu_e=0$ and
has the following explicit expression
\begin{eqnarray}
\frac{\partial V}{\partial
\mu_e}&=&\frac{1}{3\pi^2}\left[2\theta(\mu_u-m)(\mu_u^2-m^2)^{3/2}\right.\nonumber\\
&&\left.-\theta(\mu_d-m)(\mu_d^2-m^2)^{3/2}-\mu_e^3\right]=0,\nonumber\\&&\mu_d=\mu_u+\mu_e.
\end{eqnarray}
Eq.(7) is a restraint condition about $m$, $\mu_u$ and $\mu_e$. In
fact, in the effective potential (6), instead of $\mu$ and
$\mu_e$, we can first consider $\mu_u$ and $\mu_e$ as two chemical
potential variables, then by Eq.(7), only one of them, the
u-quark's chemical potential $\mu_u$, is left as a single
independent one, since $\mu_e$ may be viewed as a function of
$\mu_u$ and $m$ by Eq.(7).  This treatment will bring about great
convenience for the discussions of phase transitions.\\
\indent For research in ground state of the system, we must
consider the extreme value points of $V(m,\mu_u,\mu_e)$ determined
by the equation
\begin{widetext}
\begin{eqnarray}
\frac{\partial V}{\partial m}&=&\frac{3}{\pi^2}m\left\{\left[
\Lambda\sqrt{\Lambda^2+m_0^2}-m_0^2\ln\frac{\Lambda+\sqrt{\Lambda^2+m_0^2}}{m_0}
-(m_0\rightarrow m)\right]\right. \nonumber \\
   &&+\frac{1}{2}\left[\theta(\mu_u-m)\left(
   \mu_u\sqrt{\mu_u^2-m^2}-m^2\ln\frac{\mu_u+\sqrt{\mu_u^2-m^2}}{m}
   +(\mu_u\rightarrow \mu_d)\right]\right\}  = 0
\end{eqnarray}
and second derivation of $V(m,\mu_u,\mu_e)$ over $m$ under the
constraint given by Eq.(7)
\begin{equation}
\frac{d^2V}{dm^2}=\frac{\partial^2V}{\partial
m^2}-\left(\frac{\partial^2V}{\partial m\partial
\mu_e}\right)^2/\frac{\partial^2V}{\partial \mu_e^2}
\end{equation}
with
\begin{eqnarray}
\frac{\partial^2V}{\partial m^2}&=&\frac{3}{\pi^2}\left\{
\Lambda\sqrt{\Lambda^2+m_0^2}-m_0^2\ln\frac{\Lambda+\sqrt{\Lambda^2+m_0^2}}{m_0}
+3m^2\left(\ln\frac{\Lambda+\sqrt{\Lambda^2+m^2}}{m}-\frac{\Lambda}{\sqrt{\Lambda^2+m^2}}
\right)-\frac{\Lambda^3}{\sqrt{\Lambda^2+m^2}}
\right. \nonumber \\
&&\left.+\frac{1}{2}\left[\theta(\mu_u-m)\left(\mu_u\sqrt{\mu_u^2-m^2}
+3m^2\ln\frac{m}{\mu_u+\sqrt{\mu_u^2-m^2}} \right)
+(\mu_u\rightarrow\mu_d)\right] \right\}
\end{eqnarray}
\end{widetext}
and $\partial^2V/\partial m\partial \mu_e$ and $\partial^2V/\partial
\mu_e^2$ can be obtained from $\partial
V/\partial \mu_e$ in Eq.(7) when Eq.(3) is taken into account.\\
\indent Extremal feature of the point $m=0$ is quite important for
determination of behavior of the effective potential $V$.
Substituting $m=0$, which is obviously a solution of the extreme
value equation (8),  into Eq.(7), we will obtain the electrical
neutrality condition at $m=0$
\begin{equation}
2\mu_u^3-(\mu_u+\mu_e)^3-\mu_e^3=0.
\end{equation}
Let $\eta=\mu_e/\mu_u$, then from Eq.(11) we may obtain a real
number solution $\eta=0.256$. Furthermore, from Eqs.(9) and (10),
second derivative of $V$ over $m$ at $m=0$ becomes
\begin{equation}
\left.\frac{d^2V}{dm^2}\right|_{m=0}=\left.\frac{\partial^2V}{\partial
m^2}\right|_{m=0}=\frac{3}{2\pi^2}[1+(1+\eta)^2](\mu_u^2-\mu_{uc}^2)
\end{equation}
where
\begin{eqnarray}
\mu_{uc}^2&\equiv&\frac{2}{1+(1+\eta)^2}\left(
m_0^2\ln\frac{\Lambda+\sqrt{\Lambda^2+m_0^2}}{m_0}\right.\nonumber\\
&&\left.
-\Lambda\sqrt{\Lambda^2+m_0^2}+\Lambda^2\right).
\end{eqnarray}
Hence, $m=0$ will be a maximum (minimum) point of V if
$\mu_u^2<\mu_{uc}^2$ ($\mu_u^2>\mu_{uc}^2$), and the extremal
feature of $m=0$ will be determined by derivatives of higher order
of V over $m$ if $\mu_u^2=\mu_{uc}^2$.
\section{Second order phase transition in small
$m_0/\Lambda$}\label{small} It is easy to verify that when
$\mu_u=\mu_e=0$ (or equivalently, $\mu=\mu_e=0$), the effective
potential given by Eq.(6) will reproduce the spontaneous chiral
symmetry breaking in vacuum and the quarks get the dynamical mass
$m=m_0$. Starting from this, we will first analysis possible
chiral symmetry restoring by second order phase transition as the
chemical potential $\mu_u$ increases and derive the $\mu_u-m_0$
critical curve of the phase transition. The variation of
$V(m,\mu_u,\mu_e)$ as increase of $\mu_u$  will be discussed
successively.\\
1) $0< \mu_u < m_0 < \mu_{uc}$. Here we impose the limitation
$m_0<\mu_{uc}$ which, by Eq.(13), implies that
$\Lambda/m_0>2.865$, i.e. we are confined to the region
$m_0/\Lambda<0.349$. In this case, from Eqs.(7)-(12), the
effective potential $V$ will have the minimum point $m=m_0$ with
$\mu_e=0$ which comes from the electrical neutrality condition for
$\mu_u<m$, and the maximum point $m=0$ with the electrical
neutrality constraint $\mu_e=\eta \mu_u$ for $\mu_u>m$. A
following question is that, for the case of $\mu_u>m$, whether
there is any other non-zero solution satisfying the electrical
neutrality condition (7) for the gap equation $(\partial
V/\partial m)/m=0$ coming from Eq.(8) ? To answer this question,
we note that in this case the above two equations may be reduced
to
\begin{equation}
2(\mu_u^2-m^2)^{3/2}-(\mu_d^2-m^2)^{3/2}=\mu_e^3
\end{equation}
and
\begin{eqnarray}
&&m_0^2\ln\frac{\Lambda+\sqrt{\Lambda^2+m_0^2}}{m_0}-\Lambda\sqrt{\Lambda^2+m_0^2}=\nonumber\\
&&\frac{1}{2}m^2\ln\frac{(\Lambda+\sqrt{\Lambda^2+m^2})^2}
{(\mu_u+\sqrt{\mu_u^2-m^2})(\mu_d+\sqrt{\mu_d^2-m^2})}
\nonumber \\
&&+\frac{1}{2}(\mu_u\sqrt{\mu_u^2-m^2}+\mu_d\sqrt{\mu_d^2-m^2})
-\Lambda\sqrt{\Lambda^2+m^2}.\nonumber\\
\end{eqnarray}
\begin{widetext}
\noindent Denote that
$$
a=\Lambda/m_0,\; x=\mu_u/m_0,\; \beta=m/\mu_u,\;
\alpha=\mu_e/\mu_u
$$
then Eqs.(14) and (15) can be changed into
\begin{equation}
2(1-\beta^2)^{3/2}-[(1+\alpha)^2-\beta^2]^{3/2}=\alpha^3
\end{equation}
and
\begin{eqnarray}
 \ln(a+\sqrt{a^2+1})-a\sqrt{a^2+1}&=&
 \frac{x^2\beta^2}{2}\ln\frac{(a+\sqrt{a^2+x^2\beta^2})^2}
 {x^2(1+\sqrt{1-\beta^2})[1+\alpha+\sqrt{(1+\alpha)^2-\beta^2}]} \nonumber \\
 &&+\frac{x^2}{2}[\sqrt{1-\beta^2}+(1+\alpha)\sqrt{(1+\alpha)^2-\beta^2}]
 -a\sqrt{a^2+x^2\beta^2}
\end{eqnarray}
\end{widetext}
It is easy to check that when $a\geq 2.865$, Eqs.(16) and (17)
have no solution with $x<1$ and $\beta < 1$.  In other words, when
$a^{-1}=m_0/\Lambda\leq 0.349$, the gap equation $(\partial
V/\partial m)/m=0$ and the electrical neutrality equation (7) have
no solution with $\mu_u<m_0$ and $m<\mu_u$ indeed. \\
\indent To sum up, when $\mu_u<m_0<\mu_{uc}$, the effective
potential will have only a maximum point $m=0$ and a minimum point
$m=m_0$ accompanied with $\mu_e=0$. The latter corresponds to
ground state of the system which is similar to the case of vacuum.
This shows that spontaneous chiral symmetry breaking in vacuum
will be maintained in the case of $\mu_u<m_0$ and $\mu_e=0$. The
dynamical quark mass $m_0$ could be changed only if $\mu_u>m_0$.\\
2) $m_0<\mu_u<\mu_{uc}$. In this case, $m=0$ is still a maximum
point of $V$ by Eq.(12). On the other hand, it is not difficult to
see that now Eqs.(7) and (8) have no solution with $\mu_u<m$. Thus
we are left only the case of $\mu_u>m$ and the gap equation
$(\partial V/\partial m)/m=0$ and the electrical neutrality
condition will take the forms of Eqs.(14) and (15). In view of the
definition of $\mu_{uc}^2$ given by Eq.(13) and the constraint
$\mu_u^2<\mu_{uc}^2$, it may be deduced that Eqs.(14) and (15)
have non-zero solution $(m_1,\mu_{e1})$. From Eq.(9) we obtain
that
\begin{equation}
\left.\frac{d^2V}{dm^2}\right|_{(m_1,\mu_{e1})}=\left.\frac{\partial^2V}{\partial
m^2}\right|_{(m_1,\mu_{e1})}-\left.\left(\frac{\partial^2V}{\partial
m\partial \mu_e}\right)^2/\frac{\partial^2V}{\partial
\mu_e^2}\right|_{(m_1,\mu_{e1})}.
\end{equation}
The second term in the right-handed side of Eq.(18) is positive
since $\partial^2V/\partial\mu_e^2$ derived by Eq.(7) is always
negative. The first term may be calculated with the result
\begin{widetext}
\begin{eqnarray*}
\left.\frac{\partial^2V}{\partial m^2}\right|_{(m_1,\mu_{e1})} &=&
\left.\frac{6}{\pi^2}\left\{
\frac{m_1^2}{2}\ln\frac{(\Lambda+\sqrt{\Lambda^2+m_1^2})^2}
{(\mu_u+\sqrt{\mu_u^2-m_1^2})(\mu_d+\sqrt{\mu_d^2-m_1^2})}
-\frac{m_1^2\Lambda}{\sqrt{\Lambda^2+m_1^2}}
\right\} \right|_{\mu_{e1}}\nonumber \\
&=& \left.\frac{6}{\pi^2}\left\{
m_0^2\ln\frac{\Lambda+\sqrt{\Lambda^2+m_0^2}}{m_0}
-\Lambda\sqrt{\Lambda^2+m_0^2}+\Lambda\sqrt{\Lambda^2+m_1^2}
-\frac{1}{2}(\mu_u\sqrt{\mu_u^2-m_1^2}+\mu_d\sqrt{\mu_d^2-m_1^2})
\right.\right.\nonumber\\
&& \left.\left.-\frac{m_1^2\Lambda}{\sqrt{\Lambda^2+m_1^2}}
\right\}\right|_{\mu_{e1}}\nonumber \\
&>&\left.\frac{6}{\pi^2}\left\{
\frac{1}{2}(\mu_u^2+\mu_d^2)-\frac{1}{2}
(\mu_u\sqrt{\mu_u^2-m_1^2}+\mu_d\sqrt{\mu_d^2-m_1^2})-\Lambda^2\left(
1-\frac{\Lambda}{\sqrt{\Lambda^2+m_1^2}} \right)
\right\}\right|_{\mu_{e1}}>0,
\end{eqnarray*}
\end{widetext}
where we have used Eq.(15) with $m$ replaced by $m_1$ and the
condition $\mu_u^2<\mu_{uc}^2$, or equivalently,
$$
\frac{1}{2}[1+(1+\eta)^2]\mu_{uc}^2>\frac{1}{2}[1+(1+\eta)^2]\mu_u^2
>\left.\frac{1}{2}(\mu_u^2+\mu_d^2)\right|_{\mu_{e1}}.
$$
In this way, it is proven that
$$d^2V/dm^2|_{(m_1,\mu_{e1})}>0.$$
Since for a fixed $\mu_u$, $(m_1,\mu_{e1})$ is now the only
minimum point of $V$, so it will correspond to the ground state of
the system satisfying electrical neutrality.\\
\indent It may be found by examining Eq.(15) that when $\mu_u=m_0$
and $\mu_e=0$, we have $m=m_0$. As $\mu_u$ increases from $m_0$
and when $m_0<\mu_u<\mu_{uc}$ and $\mu_e>0$, the left-handed side
of Eq.(15) keeps unchanged, and in the right-handed side, the
second and the third term will change from zero to positive, then
the first term has to decrease so that the original $m_0$ will
change to $m<m_0$, and going up of the fourth term
$-\Lambda\sqrt{\Lambda^2+m^2}$ is consistent with the reduction of
$m$ in the first term. This means that $m$ will decrease from
$m_0$ and finally it may continuously reduce to zero, thus we come
to a critical point of second order phase transition at which the
broken chiral symmetry will be restored. The second order
$\mu_u-m_0$ critical curve will be denoted by $C_2(m_0)$ whose
equation can be obtained by setting $m=0$ in Eqs.(14) and (15) and
has the explicit expression
\begin{widetext}
\begin{equation}
\mu_u=\mu_{uc}=\left\{\frac{2}{1+(1+\eta)^2}\left[
m_0^2\ln\frac{\Lambda+\sqrt{\Lambda^2+m_0^2}}{m_0}
-\Lambda\sqrt{\Lambda^2+m_0^2}+\Lambda^2\right]\right\}^{1/2}=C_2(m_0).
\end{equation}
\end{widetext}
3) $m_0<\mu_u=\mu_{uc}$. In this case, the extreme value equation
(8) has the only solution $m=0$ and the electrical neutrality
condition (7) gives $\mu_e=\eta \mu_u$ with $\eta=0.256$. At the
only extreme point $m=0$ of $V$, the n-th derivative of $V$ over
$m$ under the electrical neutrality condition expressed by Eq.(14)
may be found out to be
\begin{equation}
\left.\dfrac{d^n V}{dm^n} \right|_{m=0}=\left\{
\begin{array}{ll}
  0, & \mathrm{when}\; n=2,3,5 \\
  \dfrac{9}{\pi^2}\ln\dfrac{\Lambda^2}{\mu_{uc}^2G(\eta)},
  & \mathrm{when}\; n=4 \\
  \dfrac{15}{\pi^2}\left(\dfrac{9}{\Lambda^2}+\dfrac{1.744}{\mu_{uc}^2}\right),
   & \mathrm{when}\;n=6 \\
\end{array}\right.
\end{equation}
where we have used the denotation
$$G(\eta)=(1+\eta)\exp\left(\frac{9+6\eta+7\eta^2}{5+2\eta+4\eta^2}\right).
$$
Eq.(20) implies that when
\begin{equation}
\mu_{uc}^2\leq\Lambda^2G^{-1}(\eta),
\end{equation}
$m=0$ will be the only minimum point of $V$ and the broken chiral
symmetry will be restored in a second order phase transition. In
view of Eq.(19), the condition (21) also means the constraint
\begin{equation}
\frac{m_0}{\Lambda}\leq 0.342,
\end{equation}
i.e. in the region $0<m_0/\Lambda\leq 0.342$, $\mu_u=\mu_{uc}$ is
a curve of second order phase transition. It is noted that the
constraint (22) is consistent with the presupposition
$m_0<\mu_{uc}$, i.e. $m_0/\Lambda <0.349$. \\
4) $\mu_u>\mu_{uc}$. We still confine ourselves to the case of
$\mu_{uc}^2\leq\Lambda^2G^{-1}(\eta)$. Obviously, by Eq.(12) and
$\mu_u>\mu_{uc}$, $m=0$ is now a minimum point of $V$. But one can
raise such a question that in the above condition, whether
$\partial V/\partial\mu_e=0$ and $(\partial V/\partial m)/m=0$
also have some non-zero $m$ solutions with $m<\mu_u$ ? To answer
this query, we can rewrite the above equations, i.e. Eqs.(14) and
(15) by
\begin{equation}
2(1-\beta^2)^{3/2}-[(1+\alpha)^2-\beta^2]^{3/2}=\alpha^3
\end{equation}
and
\begin{widetext}
\begin{eqnarray}
\frac{1}{2}[1+(1+\eta)^2]&=&\frac{\beta^2\gamma^2}{2}
\ln\frac{(b+\sqrt{b^2+\beta^2\gamma^2})^2}
{\gamma^2(1+\sqrt{1-\beta^2})(1+\alpha+\sqrt{(1+\alpha)^2-\beta^2})}
 \nonumber \\
 &&+\frac{\gamma^2}{2}\left[\sqrt{1-\beta^2}+(1+\alpha)\sqrt{(1+\alpha)^2-\beta^2}\right]
 +b^2-b\sqrt{b^2+\beta^2\gamma^2},
\end{eqnarray}
where we have used the denotations
\begin{eqnarray}
 \alpha&=&\mu_e/\mu_u,\;\beta=m/\mu_u,\; \gamma=\mu_u/\mu_{uc}\nonumber \\
  b^2 &=&
  \Lambda^2/\mu_{uc}^2=\left[1+(1+\eta)^2\right]a^2/
  2\left[\ln(a+\sqrt{a^2+1})-a\sqrt{a^2+1}+a^2\right],\;
  a=\Lambda/m_0.
\end{eqnarray}

It turns out by numerical calculation that Eqs.(23) and (24) may
have solutions with $\beta<1 \; (m<\mu_u)$ and $\gamma>1\;
(\mu_u>\mu_c)$ except that $a=\Lambda/m_0$ is quite large (e.g.
$a\geq 5$).  In other words, when $\mu>\mu_{uc}$, the effective
potential $V$ could have extreme value point with $m\neq 0$.
However, it can be proven that $m=0$ is always the least
\begin{figure*}[htb]
\includegraphics[width=16cm]{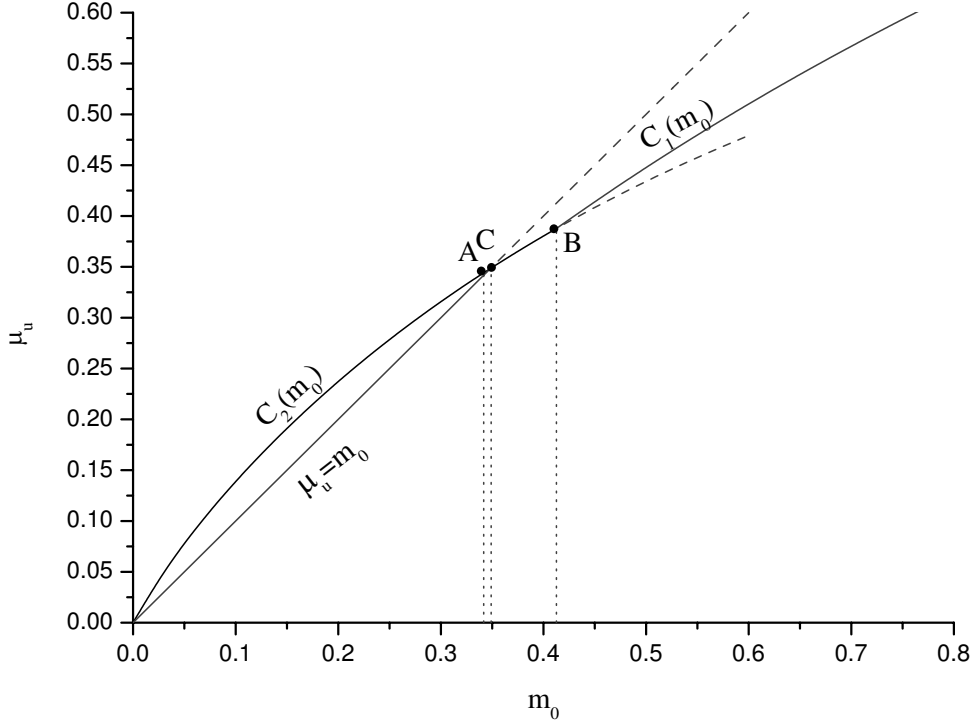}\vspace{-0.5cm}
\caption{\label{fig1}The $\mu_u-m_0$ phase diagram of the model.
Here $\mu_u$ and $m_0$ are both scaled by the momentum cutoff
$\Lambda$. The straight line through the origin and the point $C$ is
$\mu_u=m_0$. The critical curve $\mu_u=C_2(m_0)$ of second order
phase transitions starts from the origin, through the points $A$ and
$C$, ends at the point $B$. The critical curve $\mu_u=C_1(m_0)$ of
first order phase transition begins from the point $B$ then extends
to the right in the region above $\mu=\mu_{uc}$. The $B$ is a
tricritical point.}
\end{figure*}
\vspace{1cm} minimum point. In fact, by a direct calculation, we may
obtain the difference between the values of $V$ at the extreme value
points $m\neq 0$ and $m=0$ expressed by
\begin{eqnarray}
 V|_{m\neq 0}-V|_{m=0}&=&\frac{3}{4\pi^2}\left\{
2\Lambda^4-2\Lambda^3\sqrt{\Lambda^2+m^2}+\Lambda^2m^2
-\frac{1}{2}[1+(1+\eta)^2]\mu_{uc}^2m^2 \right\} \nonumber \\
   &&-\frac{1}{12\pi^2}\left.\left\{
3\mu_u(\mu_u^2-m^2)^{3/2}+3(\mu_u+\mu_e)[(\mu_u+\mu_e)^2-m^2]^{3/2}
+\mu_e^4\right\}\right|_{\mu_e=\alpha\mu_u} \nonumber \\
   && +\frac{1}{12\pi^2}\left.\left[
3\mu_u^4+3(\mu_u+\mu_e)^4+\mu_e^4\right]
\right|_{\mu_e=\eta\mu_u}\nonumber \\
   &\geq&\frac{\mu_{uc}^4}{4\pi^2} \left\{
3\left[2b^4-2b^3\sqrt{b^2+\beta^2\gamma^2}+b^2\beta^2\gamma^2
-\frac{1+(1+\eta)^2}{2}\beta^2\gamma^2\right]-(1-\beta^2)^{3/2}\right.\nonumber\\
   &&\left.-(1+\alpha)[(1+\alpha)^2-\beta^2]^{3/2}
   -\alpha^4/3+1+(1+\eta)^4+\eta^4/3\right\},\;\;
   \mathrm{when}\;\; \mu_u>\mu_{uc}.
\end{eqnarray}
\end{widetext}
In the condition $\mu_{uc}^2\leq \Lambda^2G^{-1}$ or $b^2\geq
G(\eta)$, substituting all the possible solutions of
$\alpha$,$\beta$ and $\gamma$ obtained from Eqs.(23) and (24) into
Eq.(26), we will always obtain that $V|_{m\neq 0}-V|_{m=0}>0.$ This
indicates that $m=0$ is indeed the least minimal value point of $V$
and when $\mu_u>\mu_{uc}$, the chiral symmetry has been restored
through a second order phase transition in the case of
$\mu_{uc}^2\leq \Lambda^2G^{-1}$. \indent FIG.~\ref{fig1} is the
complete $\mu_u-m_0$ phase diagram of the model. In this diagram,
the discussed second order phase transition above will correspond to
the segment of the curve $\mu_u=C_2(m_0)$ from the origin to the
point $A$ whose location is determined by the equality
$C_2(m_0)=\Lambda G^{-1/2}(\eta)$.\\

\newpage
\section{Second and first order phase transitions in larger
$m_0/\Lambda$}\label{large} It may be seen from Eq.(20) that, when
going along the curve $\mu_u=\mu_{uc}$ toward the region with
$\mu_{uc}^2>\Lambda^2G^{-1}(\eta)$, or equivalently, in view of
Eq.(22), $m_0/\Lambda>0.342$, one will get $m=0$ becoming a
maximum point of $V$, however when $\mu_u>\mu_{uc}$, $m=0$ is
again a minimum point. Such change of minimax property of $m=0$
could lead to two possibilities: either a second order phase
transition will continue or a first order phase transition will
happen. For examining a concrete realization of the above two
possibilities, we will start from the equations to determine the
critical curve of a first order phase transition. In electrical
neutrality condition, these equations read
\begin{eqnarray}
&&V(m=0)=V(m=m_1), \left.\frac{\partial
V}{\partial\mu_e}\right|_{m=0}=0,
 \left.\frac{\partial V}{\partial \mu_e}\right|_{m=m_1}= 0,\nonumber\\
&&\left.\frac{\partial V}{\partial m}/m\right|_{m=m_1}=0,
 \left.\frac{d^2V}{dm^2}\right|_{m=0}>0.
\end{eqnarray}
Since $\partial V/\partial \mu_e|_{m=0}=0$ determines only the
ratio $\eta=\mu_e/\mu_{uc}$ at $m=0$, Eqs.(27) will have the
following explicit expressions:
\begin{widetext}
\begin{eqnarray}
\frac{\mu_u^4}{3}\left[1+(1+\eta)^4+\eta^4/3\right]&=&
m_1^2\left[\frac{1+(1+\eta)^2}{2}\mu_{uc}^2 -\Lambda^2\right]
-2\Lambda^4+2\Lambda^3\sqrt{\Lambda^2+m_1^2} \nonumber \\
  &&+[\theta(\mu_u-m_1)\mu_u(\mu_u^2-m_1^2)^{3/2}/3+(\mu_u\rightarrow\mu_d)]+\mu_e^4/9,
\end{eqnarray}
\begin{equation}
2\theta(\mu_u-m_1)(\mu_u^2-m_1^2)^{3/2}
-\theta(\mu_d-m_1)(\mu_d^2-m_1^2)^{3/2}-\mu_e^3=0,
\end{equation}
\begin{eqnarray}
\frac{1+(1+\eta)^2}{2}\mu_{uc}^2&=&
\Lambda^2-\Lambda\sqrt{\Lambda^2+m_1^2}
+m_1^2\ln\frac{\Lambda+\sqrt{\Lambda^2+m_1^2}}{m_1}\nonumber \\
&&+\frac{1}{2}\left\{\theta(\mu_u-m_1)\left[
\mu_u\sqrt{\mu_u^2-m_1^2}
-m_1^2\ln\frac{\mu_u+\sqrt{\mu_u^2-m_1^2}}{m_1}\right]+(\mu_u\rightarrow
\mu_d) \right\}
\end{eqnarray}
\end{widetext}
and
\begin{equation} \mu_u^2>\mu^2_{uc}.
\end{equation}
We will discuss respectively the two cases of $\mu_{uc}>m_0$ and
$\mu_{uc}<m_0$ which could appear when
$\mu^2_{uc}>\Lambda^2G^{-1}(\eta)$.\\
1) $\mu_u>\mu_{uc}>m_0$. Since $\mu_{uc}>m_0$ can be satisfied only
if $m_0/\Lambda<0.349$ and in view of Eqs.(21) and (22), the
limitations $\mu_{uc}>m_0$ and $\mu_{uc}^2>\Lambda^2G^{-1}(\eta)$
will correspond to the region $0.342<m_0/\Lambda<0.349$ in the
$\mu_u-m_0$ plane, i.e. the region between the points $A$ and $C$ in
FIG.~\ref{fig1}. In the present case of $\mu_u>m_0$, the gap
equation (30) has no solution for $m_0<\mu_u<m_1$, so we need to
consider only the solution for $\mu_u>m_1$, thus all the
$\theta$-functions in Eqs.(28)-(30) may be removed. As a result,
Eq.(28) will be changed into
\begin{widetext}
\begin{eqnarray}
\frac{\gamma^4}{3}[1+(1+\eta)^4+\eta^4/3]&=&\beta^2\gamma^2
\left[\frac{1+(1+\eta)^2}{2}-b^2\right]-2b^4+2b^3\sqrt{b^2+\beta^2\gamma^2} \nonumber \\
 &&+\frac{\gamma^4}{3}\left\{(1-\beta^2)^{3/2}
 +(1+\alpha)[(1+\alpha)^2-\beta^2]^{3/2} +\alpha^4/3\right\},
\end{eqnarray}
\end{widetext}
Eqs.(29) and (30) will be separately identical to Eqs.(23) and
(24), and Eq.(31) will simply become $\gamma >1$, where we have
again used the denotations given by Eq.(25). By numerical solution
of Eqs. (23), (24) and (32), we have found that in the whole
region $0.342<m_0/\Lambda<0.349$, there is no solution with
$m_1\neq 0$. This means that in this region a first order phase
transition could not happen. In fact, the obtained solution is
$m_1=0$, which makes Eq.(32) becomes a trivial identity and
Eqs.(23) and (24) are reduced to $\mu_e/\mu_u=\eta$ and
$\mu=\mu_{uc}$. The last two equalities are precisely the equation
of the curve $C_2(m_0)$. Hence we can conclude that in this
region, the phase transition remains to
be second order.\\
2) $\mu_{uc}<\mu_u<m_0$. In FIG.~\ref{fig1} this corresponds to the
right-handed side of the point $C$. For $m<\mu_u$, it is easy to
verify that the electrical neutrality equation (23) and the gap
equation (24) have non-zero solution $m_1$ for $\mu_u>\mu_{uc}$ and
$\mu_{uc}^2>\Lambda^2G^{-1}(\eta)$ and $m=0$ is a minimum point.
This leads to the possibility to generate a first order phase
transition. For $m>\mu_u$, by solving Eqs.(29) and (30), it is
obtained that $m=m_0$ with $\mu_e=0$ is a minimum point. Hence, with
$m_1=m_0$ and $\mu_e=0$ being taken, Eq.(28) which determines first
order phase transition point may be changed into
\begin{widetext}
\begin{equation}
\mu_u=\left[\frac{3}{1+(1+\eta)^4+\eta^4/3}\left(
m_0^4\ln\frac{\Lambda+\sqrt{\Lambda^2+m_0^2}}{m_0}
-m_0^2\Lambda\sqrt{\Lambda^2+m_0^2}-2\Lambda^4
+2\Lambda^3\sqrt{\Lambda^2+m_0^2}\right)\right]^{1/4}\equiv
C_1(m_0).
\end{equation}
\end{widetext}
Eq.(33) expresses the equation of the curve $C_1(m_0)$. $C_1(m_0)$
may becomes a first order phase transition curve only if
$\mu_u=C_1(m_0)>\mu_{uc}$. From this constraint we obtain
\begin{equation}
m_0/\Lambda\geq 0.413
\end{equation}
which is obviously in the right-handed side of the point $C$. At
$m_0/\Lambda= 0.413$, the curve $\mu_u=C_1(m_0)$ and the curve
$\mu_u=C_2(m_0)$ intersects. The intersection point denoted by $B$
becomes the starting point of the curve $\mu_u=C_1(m_0)$
above $\mu>\mu_{uc}(m_0)$.\\
\indent In the region $0.349<m_0/\Lambda<0.413$, i.e. in the
segment between the points $C$ and $B$, similar to the case of
$\mu_{uc}>m_0$, it can be proven that Eqs.(23), (24) and (32) have
only the solution $m_1=0$, $\mu_e/\mu_u=\eta$ and $\mu=\mu_{uc}$,
i.e. the solutions are actually reduced to the second order phase
transition curve $\mu_u=C_2(m_0)$. Therefore, in the $C-B$
segment, we still have second order phase transition represented
by the critical curve $\mu_u=C_2(m_0)$.\\
\indent In summary, in the region with
$\mu_{uc}^2>\Lambda^2G^{-1}(\eta)$ or $m_0/\Lambda>0.342$, the
critical curve $\mu_u=\mu_{uc}=C_2(m_0)$ of second order phase
transition may be extended to the point $B$ where
$m_0/\Lambda=0.413$, then a critical curve $\mu_u=C_1(m_0)$ of
first order phase transition will start from the point $B$ in the
region of $\mu_u>\mu_{uc}$. So the point $B$ is a tricritical
point.\\
\indent It should be indicated that in the case without and with
electrical neutrality constraint, the feature of phase transition
of the NJL model is different. Without electrical neutrality
condition, as was discussed in Ref.[4], the second order phase
transition curve will end at a similar  point $A$, then from $A$
through the point $C$ straight to the right-handed side of the
point $B$, one will always have a first order phase transition
curve. However, in present case with electrical neutrality
requirement, from $A$ to $B$, one continue to get a second order
phase
transition, instead of a first order one.\\
\indent The total conclusions of this paper come from a general
research of the used NJL model, i.e. the parameters of the model,
the 3D momentum cutoff $\Lambda$ and the dynamical quark mass
$m_0$ in vacuum (correspondingly, the four-fermion coupling
constant $G_S$) have been considered as arbitrary ones. If the
discussed NJL model is used to simulate QCD for normal quark
matter and the conventional phenomenological values of the
parameters \cite{kn:5}
$$
G_S=5.0163 GeV^{-2}, \Lambda=0.6533 GeV
$$
are taken, then by Eq.(4) we will obtain $m_0/\Lambda=0.48$. Based
on the results of present paper, chiral symmetry restoring at high
density in this model must be first order phase transition.
\section{Conclusions}\label{conc}
In this paper, we have generally analyzed chiral symmetry
restoring phase transitions at zero temperature and high density
in a NJL model to describe two-flavor normal quark matter under
electrical neutrality condition. It has been found that the
feature of phase transitions is decided by the ratio
$m_0/\Lambda$, where $m_0$ is the dynamical quark mass in vacuum
and $\Lambda$ is the 3D momentum cutoff of the loop integrals.
Depending on $m_0/\Lambda$ is less or greater than 0.413, the
phase transition will be second or first order. As a comparison,
the resulting region where a second order phase transition happens
is wider than the one in the case without electrical neutrality
constraint. For the value of $m_0/\Lambda$ based on QCD
phenomenology, the phase transition must be first order. \\
\indent The present discussions of normal quark matter based on a
NJL model can be generalized to the case of color superconducting
quark matter where one must also consider diquark condensates,
besides the quark-antiquark condensates, and at the same time,
also impose color neutrality condition, besides electrical
neutrality one.\\

\end{document}